\documentclass[secnumarabic,amssymb,nobibnotes,aps,prc]{revtex4}

\usepackage{epsfig}
\usepackage{graphicx}
\usepackage{amsmath}
\usepackage{hyperref}

\begin{document} 

\title{Coherent photoproduction of vector mesons in heavy ion ultraperipheral collisions: Update for  
run 2 at the CERN Large Hadron Collider}

\author{V. Guzey}
\affiliation{Petersburg Nuclear Physics Institute (PNPI), National Research Center ``Kurchatov Institute'', Gatchina, 188300, Russia}
\author{E. Kryshen}
\affiliation{Petersburg Nuclear Physics Institute (PNPI), National Research Center ``Kurchatov Institute'', Gatchina, 188300, Russia}
\author{M. Zhalov}
\affiliation{Petersburg Nuclear Physics Institute (PNPI), National Research Center ``Kurchatov Institute'', Gatchina, 188300, Russia}


\begin{abstract}

We make predictions for the cross sections of coherent photoproduction of $\rho$, $\phi$, $J/\psi$, $\psi(2S)$, and $\Upsilon(1S)$ mesons in Pb-Pb ultraperipheral collisions (UPCs) at $\sqrt{s_{NN}}=5.02$ TeV in the kinematics of run 2 at the Large
Hadron Collider extending
the approaches successfully describing the available Pb-Pb UPC data at $\sqrt{s_{NN}}=2.76$ TeV. 
Our results illustrate the important roles of hadronic fluctuations of the photon and inelastic nuclear 
shadowing in photoproduction of light vector mesons on nuclei and the large leading twist nuclear gluon shadowing
in photoproduction of quarkonia on nuclei.
We show that the ratio of $\psi(2S)$ and $J/\psi$ photoproduction cross sections in Pb-Pb UPCs
is largely determined by the ratio of these cross sections on the proton.
We also argue that UPCs with electromagnetic excitations of the colliding ions
followed by the forward neutron emission allows one to significantly increase the range of photon energies accessed 
in vector meson photoproduction on nuclei.

\end{abstract}

\maketitle

\section{Introduction}
\label{sec:intro}

Ultrarelativistic collisions of ions at large transverse distances (impact parameters) exceeding the sum of their radii
are called ultraperipheral collisions (UPCs). For such collisions the strong interaction is suppressed and the ions 
interact through the emission of quasi-real photons. 
The flux of these photons has the high intensity and the wide energy spectrum
allowing one to study photon-induced processes at unprecedentedly high energies~\cite{Baltz:2007kq}.
In particular, UPCs at the Large Hadron Collider (LHC) provide the photon energies exceeding those achieved 
in lepton--proton scattering at the Hadron-Electron Ring Accelerator (HERA) and in nucleus--nucleus UPCs at the
Relativistic Heavy Ion Collider (RHIC) by at least a factor of 10.

The UPC program at the LHC during run 1 focused primarily on exclusive photoproduction of 
light and heavy vector mesons. 
Coherent photoproduction of $\rho$ mesons in nucleus--nucleus ($AA$) UPCs was measured by the ALICE Collaboration 
at $\sqrt{s_{NN}}=2.76$ TeV~\cite{Adam:2015gsa} continuing and extending to higher energies the measurement of this 
process by the STAR Collaboration at RHIC at $\sqrt{s_{NN}}=64.4$, 130 and 200 GeV~\cite{Adler:2002sc,Abelev:2007nb,Agakishiev:2011me} ($\sqrt{s_{NN}}$ is the invariant collision energy per nucleon).
These data probe the dynamics of soft high-energy $\gamma p$ and $\gamma A$ interactions and test models 
of nuclear shadowing in photon--nucleus scattering, see, e.g.~\cite{Frankfurt:2015cwa}.

Turning to heavy vector mesons, exclusive photoproduction of charmonia --- $J/\psi$ and $\psi(2S)$ mesons --- was 
measured in proton--proton ($pp$) UPCs at
$\sqrt{s_{NN}}=7$ TeV by the LHCb Collaboration~\cite{Aaij:2013jxj,Aaij:2014iea},
 in proton--nucleus ($pA$) UPCs at $\sqrt{s_{NN}}=5.02$ TeV by the ALICE Collaboration~\cite{TheALICE:2014dwa}, and in
 Pb-Pb UPCs at $\sqrt{s_{NN}}=2.76$ TeV by the ALICE Collaboration~\cite{Abbas:2013oua,Abelev:2012ba,Adam:2015sia}.
 Also, exclusive photoproduction of bottomia --- the $\Upsilon(1S)$, $\Upsilon(2S)$ and $\Upsilon(3S)$ mesons --- in $pp$ UPCs at $\sqrt{s_{NN}}=7$ and 8 TeV 
 was recently measured by the LHCb Collaboration~\cite{Aaij:2015kea}.
These data probe the gluon distribution of the target $g(x,\mu^2)$
at small values of the momentum fraction $x$ and the resolution scale $\mu^2={\cal O}(\rm few)$ GeV$^2$ ($J/\psi$)~\cite{Ryskin:1992ui} and $\mu^2={\cal O}(\rm few~tens)$ GeV$^2$ ($\Upsilon$).
In particular, 
 the analyses of these data at leading-order (LO) and next-to-leading order (NLO) QCD have provided new constraints on 
 the small-$x$ behavior of the gluon distribution in the proton 
$g_p(x,\mu^2)$ down to $x=6 \times 10^{-6}$~\cite{Jones:2013pga,Guzey:2013qza}
and the gluon distribution in heavy nuclei $g_A(x,\mu^2)$ down to $x \approx 10^{-3}$~\cite{Adeluyi:2012ph,Guzey:2013xba}.
The data also impose constraints on the parameters and the strong interaction dynamics of the color 
dipole model approach~\cite{Goncalves:2004bp,Lappi:2013am,Goncalves:2014wna} 
and the STARlight Monte Carlo generator~\cite{Klein:1999qj}.

Using the phenomenological frameworks allowing one to successfully describe exclusive 
photoproduction of $\rho$ and $J/\psi$ mesons in Pb-Pb UPCs at the LHC at $\sqrt{s_{NN}}=2.76$ TeV 
and extending them to the cases of $\phi$ and $\Upsilon$ mesons, respectively, we calculate the cross sections
of coherent photoproduction of $\rho$, $\phi$, $J/\psi$, $\psi(2S)$, and $\Upsilon(1S)$ mesons in Pb-Pb UPCs 
at $\sqrt{s_{NN}}=5.02$ TeV in the kinematics of run 2 at the LHC. 
The aim is to present the baseline predictions of the approaches based on the combination of the modified
vector meson dominance (mVMD) model with the Gribov--Glauber (GG) model  of nuclear shadowing for photoproduction of 
$\rho$ and $\phi$ mesons and perturbative QCD with the leading twist gluon 
nuclear shadowing for photoproduction of $J/\psi$, $\psi(2S)$, and 
$\Upsilon(1S)$ mesons. A comparison of these predictions with the run 2 UPC data will test the role of inelastic nuclear shadowing in exclusive photoproduction of $\rho$ and $\phi$ mesons on nuclei, constrain the effect of leading twist nuclear shadowing
in the gluon distribution $g_A(x,\mu^2)$ of heavy nuclei down to $x \approx 5 \times 10^{-4}$ for $\mu^2={\cal O}(2-20)$ GeV$^2$,
and address the issue of the nuclear suppression of the cross section of $\psi(2S)$ photoproduction on Pb, 
whose measurement in run 1 disagrees with most of theoretical expectations~\cite{Adam:2015sia}.

\section{Coherent photoproduction of vector mesons in UPCs}
\label{sec:formalism}

The expression for the cross section of exclusive photoproduction of vector mesons in ion UPCs is well known and reads, see, e.g., Ref.~\cite{Baltz:2007kq},
\begin{equation}
\frac{d \sigma_{AA \to V AA}(y)}{dy}=N_{\gamma/A}(y) \sigma_{\gamma A \to V A}(y)+N_{\gamma/A}(-y) \sigma_{\gamma A \to V A}(-y) \,,
\label{eq:cs_upc}
\end{equation}
where $y$ is the rapidity of the vector meson $V$ ($V=\rho, \phi, J/\psi, \psi(2S), \Upsilon(1S)$);
$N_{\gamma/A}(y)$ is the photon flux of ion $A$; 
$\sigma_{\gamma A \to V A}(y)$ is the
cross section of exclusive photoproduction of the vector meson $V$ on the hadronic target $A$.
The rapidity $y$ is related to the photon energy in the laboratory reference frame, $\omega$: 
$y=\ln(2 \omega /M_V)=\ln(W_{\gamma p}^2/(2 \gamma_L m_N M_V))$, where $M_V$ is the vector meson mass, 
$W_{\gamma p}$ is the invariant photon--nucleon center-of-mass energy, $\gamma_L$ is the Lorentz factor of the ion emitting
the photon, and $m_N$ is the nucleon mass. The presence of two terms in Eq.~(\ref{eq:cs_upc}) reflects the fact 
that either 
of the colliding ions can serve as a photon source and as a target. It leads to the ambiguity
in the value of $W_{\gamma p}$ common for symmetric UPCs: at given $y$, the UPC cross section in Eq.~(\ref{eq:cs_upc})
is a sum of the high-$W_{\gamma p}$ and low-$W_{\gamma p}$ contributions.

The flux $N_{\gamma/A}(\omega)$ in Eq.~(\ref{eq:cs_upc}) can be found using the well-known Weizs{\"a}cker--Williams 
approximation of quantum electrodynamics~\cite{Budnev:1974de,Bertulani:1987tz}. In addition, in 
$AA$ UPCs one needs 
 to take into account the suppression of the strong interaction 
between the colliding ions at small impact parameters. The resulting expression for $N_{\gamma/A}(\omega)$ reads:
\begin{eqnarray}
N_{\gamma/A}(\omega) = \int d^2 \vec{b} \, \Gamma_{AA}(b) N_{\gamma/A}(\omega,b) \,,
\label{eq:flux_AB}
\end{eqnarray} 
where $b$ is the impact parameter (the transverse distance between the ion centers of mass);
$\Gamma_{AA}(b)$ is the probability to not have the strong interaction between the ions at the impact parameter $b$;
 $N_{\gamma/A}(\omega,b)$ is the impact parameter dependent photon flux of ion $A$.
 
The probability $\Gamma_{AA}(b)$ is given by the standard expression of the Glauber model for high-energy nucleus--nucleus scattering~\cite{Glauber:1970jm},
\begin{equation}
\Gamma_{AA}(b)=\exp\left(-\sigma_{NN}^{\rm tot} \int d^2 \vec{b}_1 T_A(\vec{b}_1) T_A(\vec{b}-\vec{b}_1)\right) \,,
\label{eq:Gamma_AA}
\end{equation}
where $\sigma_{NN}^{\rm tot}$ is the total nucleon--nucleon cross section~\cite{Agashe:2014kda}; 
$T_A(b)=\int dz \rho_A(b,z)$ is the nuclear optical density, where 
$\rho_A$ is the density of nucleons. In our analysis, we used $\rho_A$ calculated in the Hartree--Fock--Skyrme model~\cite{Beiner:1974gc}.

The impact parameter dependent photon flux $N_{\gamma/A}(\omega,b)$ entering Eq.~(\ref{eq:flux_AB}) is given
by the following expression, see, e.g., Ref.~\cite{Vidovic:1992ik},
\begin{equation}
N_{\gamma/A}(\omega,b)=\frac{\alpha_{\rm e.m.}Z^2}{\pi^2}  \left|\int_0^{\infty}  \frac{dk_{\perp} k_{\perp}^2}{k_{\perp}^2 +
\omega^2/\gamma_L^2} F_{\rm ch}(k_{\perp}^2+\omega^2/\gamma_L^2) J_1(b k_{\perp}) \right|^2 \,,
\label{eq:flux_vidovic}
\end{equation}
where $\alpha_{\rm e.m.}$ is the fine-structure constant; $Z$ and $F_{\rm ch}(k_{\perp}^2)$ are the charge and the charge form factor
of the ion emitting the photon, respectively; 
$k_{\perp}$ is the photon  transverse momentum;
$J_1$ is the Bessel function of the first kind.

Coherent photoproduction of vector mesons in UPCs can also be accompanied by additional multiple photon exchanges between the colliding nuclei leading to electromagnetic excitations of one or both nuclei followed by forward neutron 
emission~\cite{Baltz:2002pp}. These neutrons can be detected by zero degree calorimeters (ZDCs) placed at long distances on
both sides of the detector; the corresponding classes of UPC events are distinguished by the number of neutrons detected in these ZDCs.
Assuming that all photon emissions are independent, additional photon exchanges lead to the following 
modification of the photon flux of equivalent photons: 
\begin{equation}
N_{\gamma/A}^i(\omega)=\int d^2 \vec{b} P_i(b) \Gamma_{AA}(b) N_{\gamma/A}(\omega,b) \,, 
\label{eq:N_i}
\end{equation}
where $P_i(b)$ is the probability of Coulomb excitations at the impact parameter $b$ in channel $i$;
$i=(\rm 0n0n, 0nXn,XnXn)$ labels various channels corresponding to a different number of neutrons detected in both ZDCs.
For instance, the 0n0n-channel corresponds to selection of events with no neutrons in either of ZDCs;
the 0nXn-channel corresponds to the detection of zero neutrons in one ZDC and at least one neutron in another one indicating
the Coulomb excitation of only one of the colliding nuclei; the XnXn-channel corresponds to the mutual electromagnetic
excitation of the nuclei followed by forward neutron emission when each of the ZDCs detects at least one fast neutron.

Note that the characteristic impact parameter in $N_{\gamma/A}^i(\omega)$ for the 0nXn and XnXn-channels is smaller than that in 
$N_{\gamma/A}(\omega)$~\cite{Baltz:2002pp}, which
shifts the strength of $N_{\gamma/A}^i(\omega)$ towards larger values of $\omega$ --- this effect is directly seen in
 the rapidity distributions for the 0nXn and XnXn-channels shown in Figs.~\ref{fig:rho}--\ref{fig:Upsilon}.

In Eq.~(\ref{eq:cs_upc}), the key quantity containing information on the dynamics of coherent photoproduction of 
light and heavy vector mesons and the nuclear structure is the $\sigma_{\gamma A \to V A}$ cross section. 
Below we summarize
the two frameworks that we use to calculate the cross sections of 
nuclear coherent photoproduction of light $\rho$ and $\phi$ mesons and heavy $J/\psi$, $\psi(2S)$, and $\Upsilon(1S)$ mesons, 
respectively.

\subsection{Combination of modified VMD and Gribov--Glauber models for nuclear photoproduction of light vector mesons}

At high energies, the photon interacts with protons and nuclei by means of its hadronic fluctuations.  
The lifetime of these fluctuations linearly increases with an increase of the photon energy $\omega$ and, hence, for
sufficiently large $\omega$, the photon can be represented as a coherent superposition of hadronic fluctuations, which 
in general have rather different invariant masses and cross sections of the interaction with the target.

The well-known realization of this picture of high-energy photon--hadron interactions is the vector meson dominance (VMD)
model and its extension to the generalized vector meson dominance (GVMD) model,
for review, see, e.g.~\cite{Bauer:1977iq}. 
The two-state GVMD model describes within experimental uncertainties the $\gamma A \to \rho A$ cross section 
measured in fixed-target experiments and Au-Au UPCs for $\sqrt{s_{NN}} \leq 130$ GeV at RHIC covering the 
$3.5 \ {\rm GeV} < W_{\gamma p} \leq 10$ GeV range~\cite{Frankfurt:2002wc,Frankfurt:2002sv}.
For higher photon energies relevant for Au-Au UPCs at $\sqrt{s_{NN}} = 200$ GeV at RHIC
and Pb-Pb UPCs at the LHC, one needs to include the contribution of high-mass fluctuations of the photon~\cite{Frankfurt:2015cwa}.
This contribution is required and constrained by the significant cross section of photon inelastic diffraction
into large masses and the data on diffractive photoproduction of light vector mesons on the proton at the HERA;
it leads to the sizable inelastic nuclear shadowing in the $\gamma A \to V A$ cross section, where 
$V$ denotes $\rho$ or $\phi$ mesons here.

The convenient way to phenomenologically implement these observations in the calculation of 
the $\gamma A \to \rho A$ and $\gamma A \to \phi A$ cross sections is given by the formalism of cross section fluctuations. 
In the modified VMD (mVMD) model of Ref.~\cite{Frankfurt:2015cwa}, the photon fluctuates into the vector meson $V$,
which interacts with the nuclear target as a coherent superposition of 
eigenstates of the scattering operator, whose eigenvalues are 
the scattering cross sections $\sigma$;
the weight of a given fluctuation is given by the distribution $P_{V}(\sigma)$.
Each state interacts with nucleons of the target nucleus 
according to the Gribov--Glauber (GG) model of nuclear shadowing.
The resulting $\gamma A \to V A$ amplitude is summed over all possible fluctuations, which 
corresponds to averaging with the distribution 
$P_{V}(\sigma)$. Therefore, one obtains for the $\gamma A \to \rho A$ and $\gamma A \to \phi A$ cross sections
in the high-energy limit~\cite{Frankfurt:2015cwa}:
\begin{equation}
\sigma_{\gamma A \to V A}^{\rm mVMD-GGM} = \left(\frac{e}{f_{V}}\right)^2  
\int d^2\vec{b} \left| \int d\sigma P_{V}(\sigma) \left(1-e^{-\frac{\sigma}{2}  T_A(b)} \right)\right|^2  \,, 
\label{eq:cs_rho_approx_CF}
\end{equation}
where $f_{\rho}^2/(4 \pi)=2.01\pm 0.1$ is determined from the $\rho \to e^{+} e^{-}$ decay; 
$f_{\phi}^2/(4 \pi)=13.7$~\cite{Klein:1999qj}.

\begin{figure}[t]
\begin{center}
\epsfig{file=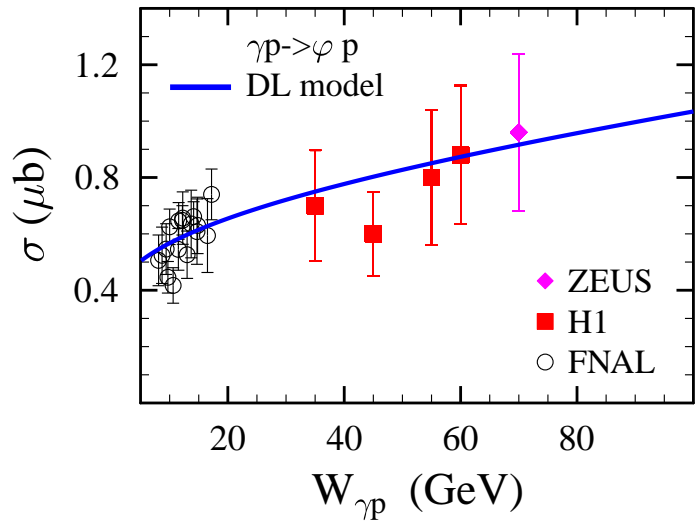,scale=0.75}
 \caption{The $\gamma p \to \phi p$ cross section as a function of $W_{\gamma p}$. The fixed-target~\cite{Egloff:1979mg,Busenitz:1989gq} 
 and HERA~\cite{Derrick:1996af,Berger_thesis} data vs.~the Donnachie--Landshoff (DL) model~\cite{Donnachie:1994zb}.}
 \label{fig:fitot}
\end{center}
\end{figure}

The distribution $P_{V}(\sigma)$ cannot be calculated from the first principles. 
Using the additive quark model,
it is natural to assume that $P_{V}(\sigma)$ 
should be similar to the pion $P_{\pi}(\sigma)$; the latter is constrained by perturbative QCD in the small-$\sigma$ limit,
the counting quark rule relation to cross section fluctuations in the proton and the data on inelastic nuclear shadowing
in pion--deuteron scattering~\cite{Blaettel:1993ah}. 
Therefore, in our calculations we use~\cite{Frankfurt:2015cwa}:
\begin{equation}
P_{V}(\sigma)=C \frac{1}{1+(\sigma/\sigma_0)^2} e^{-(\sigma/\sigma_0 -1)^2/\Omega^2} \,.
\label{eq:Psigma}
\end{equation}
The free parameters $C$, $\sigma_0$ and $\Omega$ are found from 
the following constraints:
\begin{eqnarray}
\int d\sigma P_{V}(\sigma) &=& 1 \,, \nonumber\\
\int d\sigma P_{V}(\sigma) \sigma  &=&  \langle \sigma \rangle =\hat{\sigma}_{V N} \,, \nonumber\\
\int d\sigma P_{V}(\sigma) \sigma^2  &=& \langle \sigma \rangle^2 (1+\omega_{\sigma}^{V}) \,.
\label{eq:P_pion2}
\end{eqnarray}
The effective meson--nucleon cross section $\hat{\sigma}_{V N}=(f_{V}/e)(16 \pi d\sigma_{\gamma p \to V p}(t=0)/dt)^{1/2}$
is determined from the fit to the experimental data on the forward $d\sigma_{\gamma p \to V p}(t=0)/dt$ cross section.
For the $\rho$ meson, we used the fit of Ref.~\cite{Frankfurt:2015cwa}; for the $\phi$ meson, we used 
the Donnachie--Landshoff (DL) model~\cite{Donnachie:1994zb}. 
The description of the available data on $\phi$ photoproduction on the proton in the DL model is shown in Fig.~\ref{fig:fitot}.
Note that the significant experimental uncertainties of the data shown in Fig.~\ref{fig:fitot} may affect 
our predictions for the $\gamma A \to \phi A$ cross section.

The parameter $\omega_{\sigma}^{V}$ quantifies
the dispersion of $P_{V}(\sigma)$, i.e., the strength of cross section fluctuations. For the $\rho$ meson,  
it is determined using 
experimental information on the photon diffraction dissociation~\cite{Chapin:1985mf}, see the details 
in Ref.~\cite{Frankfurt:2015cwa}.
For the $\phi$ meson, using the independence of the ratio of the high-energy diffraction dissociation
and the total cross sections from the projectile type~\cite{Cool:1981uw},
one estimates that $\omega_{\sigma}^{\phi} \approx (\sigma_{NN}/\hat{\sigma}_{\phi N}) \omega_{\sigma}^{N}$,
where $\omega_{\sigma}^{N}$ quantifies the dispersion of cross section fluctuations of the nucleon, 
which is constrained by the data on nucleon--deuteron inelastic nuclear shadowing and
diffraction dissociation in proton--antiproton scattering~\cite{Blaettel:1993ah}.

\begin{figure}[t]
\begin{center}
\epsfig{file=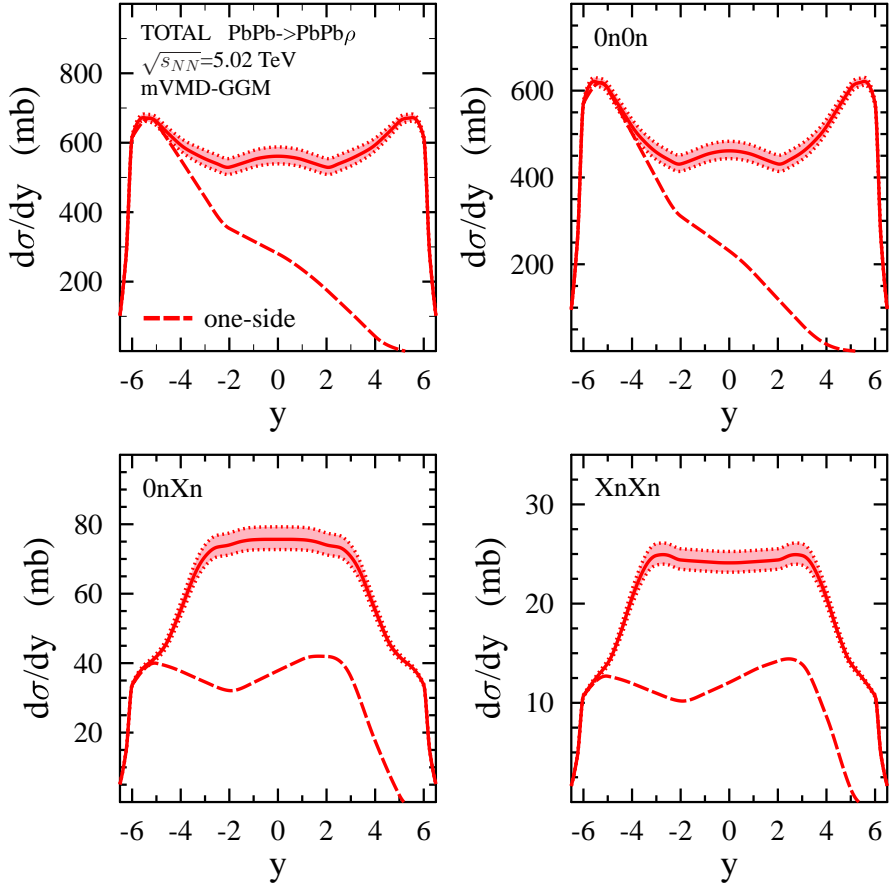,scale=0.6}
 \caption{The $d \sigma_{Pb Pb \to \rho Pb Pb}(y)/dy$ cross section as a function of the $\rho$ meson rapidity $y$
 at $\sqrt{s_{NN}}=5.02$ TeV. Predictions of the combination of the modified VMD 
 and GG models (mVMD-GGM) are shown for the four considered reaction channels.
The solid curves correspond to the calculation using the nominal value of $\omega^{\rho}_{\sigma}$;
the shaded areas show the theoretical uncertainty in modeling of this quantity.
The dashed curves labeled ``one-side" show the contribution of the first term in Eq.~(\ref{eq:cs_upc}).}
 \label{fig:rho}
\end{center}
\end{figure}

\begin{figure}[t]
\begin{center}
\epsfig{file=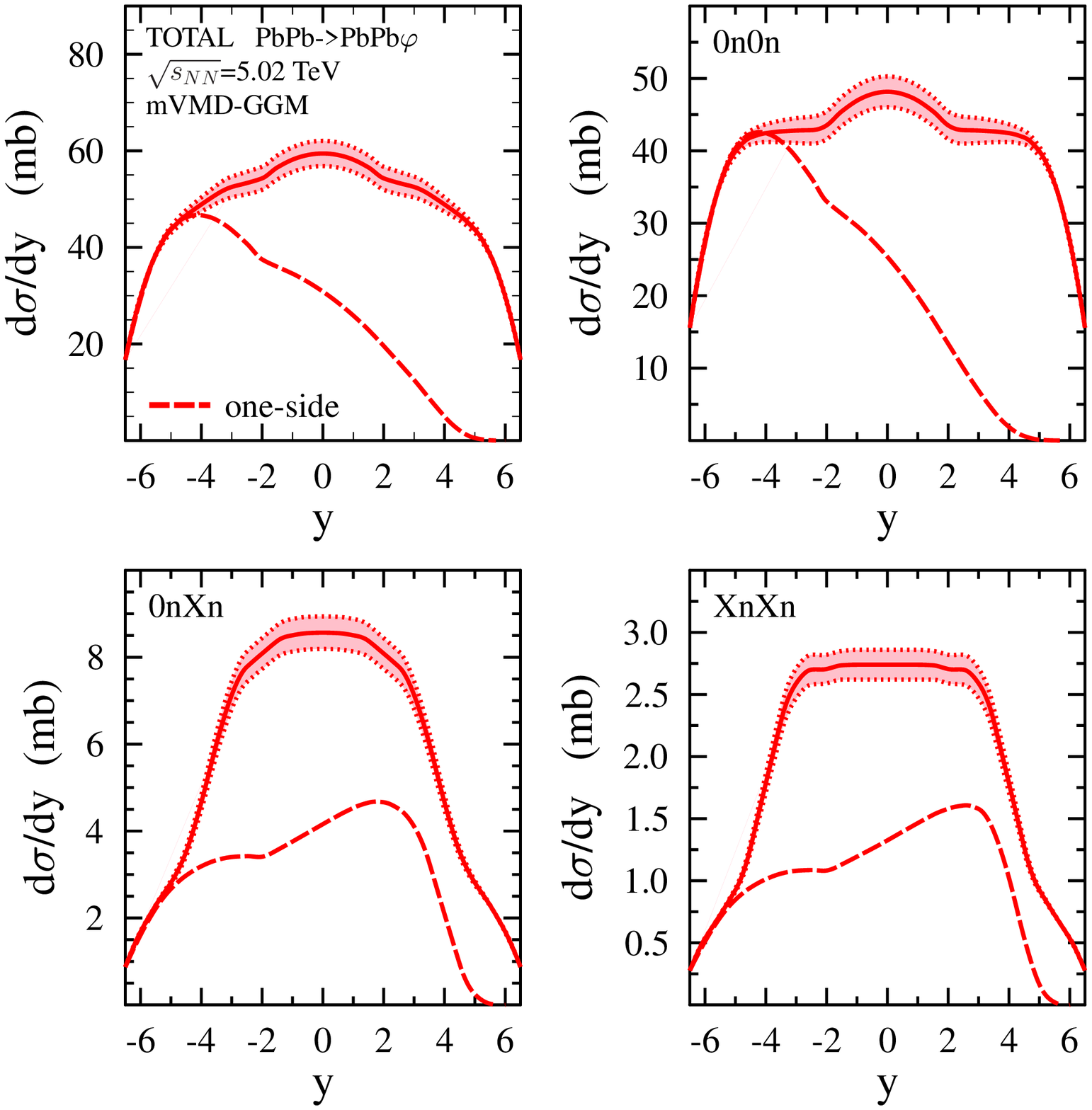,scale=0.6}
 \caption{The $d \sigma_{Pb Pb \to \phi Pb Pb}(y)/dy$ cross section as a function of the $\phi$ meson rapidity $y$
 at $\sqrt{s_{NN}}=5.02$ TeV. For notation, see Fig.~\ref{fig:rho}.
}
 \label{fig:phi}
\end{center}
\end{figure}

Our results for coherent photoproduction of $\rho$ and  $\phi$ vector mesons in
Pb-Pb UPCs at $\sqrt{s_{NN}}=5.02$ TeV at the LHC are shown in Figs.~\ref{fig:rho} and \ref{fig:phi}, respectively.
In each plot, we give predictions for the UPC cross section calculated using $N_{\gamma/A}(\omega)$~(\ref{eq:flux_AB}) in Eq.~(\ref{eq:cs_upc}) (labeled ``Total'') and UPC cross sections explicitly taking into account or prohibiting
additional Coulomb excitations of the nuclei followed by forward neutron emission, which are calculated using 
$N_{\gamma/A}^i(\omega)$~(\ref{eq:N_i}) in Eq.~(\ref{eq:cs_upc}) in three different channels $i=(\rm 0n0n, 0nXn, XnXn)$ (the curves 
and the panels are labeled accordingly). Naturally, $d\sigma({\rm total})/dy=d\sigma({\rm 0n0n})/dy+2 d\sigma({\rm 0nXn})/dy
+d\sigma({\rm XnXn})/dy$.

In each panel, we show the net result of Eq.~(\ref{eq:cs_upc}) by solid curves and the contribution
of only the first term in Eq.~(\ref{eq:cs_upc}) labeled ``one-side" by dashed curves.
They correspond to the calculation using the nominal value of $\omega^{V}_{\sigma}$; 
the shaded areas (for the net result only) show the theoretical uncertainty in modeling of this quantity,
see the details in Ref.~\cite{Frankfurt:2015cwa}.

Electromagnetic excitations of the nuclei resulting in the yield of
forward neutrons radically change the rapidity distribution: additional photon exchanges between the colliding ions select the 
region of smaller impact parameters in Eq.~(\ref{eq:N_i}) and, hence, enhance the high-$W_{\gamma p}$ contribution 
to nuclear photoproduction of vector mesons~\cite{Baltz:2002pp}. This is clearly seen from the comparison of the dashed curves
(``one-side") in the different panels of Figs.~\ref{fig:rho} and \ref{fig:phi}.

\subsection{Exclusive photoproduction of charmonia on nuclei in LO perturbative QCD}

In the leading logarithmic approximation of perturbative QCD (pQCD), the cross section of exclusive photoproduction of
charmonia, i.e., vector mesons consisting of the charm quark and antiquark in the lowest Fock state,
on the proton reads
\begin{equation}
\frac{d \sigma_{\gamma p \to V p}(W_{\gamma p},t=0)}{dt}=C_p(\mu^2) [\alpha_s(\mu^2) x g_p(x,\mu^2)]^2 \,,
\label{eq:charm}
\end{equation}
where $V$ stands for $J/\psi$ or $\psi(2S)$ mesons [$\psi(2S)$ is the first radially-excited 
charmonium state with $J^{PC}=1^{--}$]; $\alpha_s(\mu^2)$ is the strong coupling constant; $x g_p(x,\mu^2)$
is the gluon density of the proton evaluated at the light-cone momentum fraction $x=M_V^2/W_{\gamma p}^2$ and
the resolution scale $\mu$; $C_p(\mu^2)$ is the normalization factor depending on approximations used in the evaluation
of the  $\gamma p \to V p$ amplitude. 

In the case of $J/\psi$ photoproduction on the proton, Eq.~(\ref{eq:charm}) was first derived in Ref.~\cite{Ryskin:1992ui}
using the non-relativistic approximation for the $J/\psi$ wave function; it was found that $\mu^2=M_{J/\psi}^2/4=2.4$ GeV$^2$
and $C_p(\mu^2)=\pi^3 \Gamma_{ee} M_{J/\psi}^3/(48 \alpha_{\rm e.m.} \mu^8)$, where $\Gamma_{ee}$ is the $J/\psi \to e^{+} e^{-}$ decay width and $\alpha_{\rm e.m.}$ is the fine-structure constant. 
Going beyond this approximation~\cite{Ryskin:1995hz,Martin:2007sb}, one obtains $C_p(\mu^2)=F^2(\mu^2) \bar{R}_g^2 (1+\eta^2) \pi^3 \Gamma_{ee} M_{J/\psi}^3/(48 \alpha_{\rm e.m.} \mu^8)$, where $\eta$ is the ratio of the real to the imaginary parts of 
the $\gamma p \to J/\psi p$ scattering amplitude, $\bar {R}_g \approx 1.2$ is the skewness factor describing the 
enhancement of the $\gamma p \to J/\psi p$ amplitude due to its off-forward kinematics, and
$F^2(\mu^2) \approx 0.5$ is the factor taking into account
the effects of the quark transverse momentum  in the $J/\psi$ wave function. Note that Eq.~(\ref{eq:charm})
can also be generalized beyond the leading logarithmic approximation by including the gluon transverse momenta
in the gluon ladder~\cite{Ryskin:1995hz}. 

For the case of $\psi(2S)$, the same framework is immediately applicable with $\mu^2=M_{\psi(2S)}^2/4 = 3.4$ GeV$^2$
in the non-relativistic limit for the $\psi(2S)$ wave function~\cite{Jones:2013eda}.

The nonzero charm quark transverse momentum in the charmonium wave function
 leads to an effective increase of the resolution scale $\mu^2$ at which the gluon distribution in Eq.~(\ref{eq:charm}) 
 is probed.
 In our approach, we determine $\mu^2$ phenomenologically by requiring that Eq.~(\ref{eq:charm}) with a wide array of 
 modern leading-order (LO) gluon distributions of the proton describes the 
 high-$W_{\gamma p}$ dependence of the $\sigma_{\gamma p \to J/\psi p}(W_{\gamma p})$ cross section measured 
 at the HERA and 
 the LHC by the LHCb Collaboration and the $\sigma_{\gamma p \to \psi(2S) p}(W_{\gamma p})$ cross section 
 measured at the HERA.
This gives $\mu^2  \approx 3$ GeV$^2$ for $J/\psi$~\cite{Guzey:2013qza} and  $\mu^2  \approx 4$ GeV$^2$ for $\psi(2S)$~\cite{Guzey:2014axa}, 
respectively.
The factor of $C_p(\mu^2)$ is chosen to reproduce the normalization of the respective experimental cross sections at $W=100$ GeV.
The resulting LO pQCD framework based on Eq.~(\ref{eq:charm}) provides good description of all high-energy HERA and LHC data
on charmonium [$J/\psi$ and $\psi(2S)$] photoproduction on the proton.

Note that in our phenomenological approach to charmonium photoproduction on the proton, we determine the resolution scale $\mu^2$ and the cross section normalization using the high-energy HERA data, which allows us to effectively include 
sizable corrections to the non-relativistic approximation for the charmonium wave function~\cite{Ryskin:1995hz}.
In particular, our result for the $\psi(2S)$ case agrees with that obtained in the framework of the color dipole model, 
where the presence of the node in the $\psi(2S)$  wave function leads to the intricate cancellation between the short-distance and the long-distance contributions to the photoproduction cross section leading to an enhancement of the contribution of 
small-size dipoles, which decreases the average size of $\psi(2S)$ probed in high-energy photoproduction~\cite{Nemchik:1997xb,Hufner:2000jb,Ducati:2013bya}. This allows one to 
reproduce within experimental uncertainties the HERA measurements~\cite{Adloff:2002re} of the
$W_{\gamma p}$ dependence and the normalization of the 
$\sigma_{\gamma p \to \psi(2S) p}/\sigma_{\gamma p \to J/\psi p}$ cross section ratio
and the slope of the differential $d\sigma_{\gamma p \to \psi(2S) p}(t)/dt$ cross section~\cite{Nemchik:1997xb,Hufner:2000jb}.

The application of Eq.~(\ref{eq:charm}) to nuclear targets allows one to consider coherent photoproduction of charmonia 
on nuclei in pQCD. The corresponding cross section integrated over the momentum transfer $t$ reads~\cite{Guzey:2013qza}
\begin{eqnarray}
\sigma_{\gamma A \to V A}(W_{\gamma p}) &=& C_A(\mu^2) [\alpha_s(\mu^2) x g_A(x,\mu^2)]^2 \Phi_A(t_{\rm min}) \nonumber\\
&=&\frac{C_A(\mu^2)}{C_p(\mu^2)} \frac{d \sigma_{\gamma p \to V p}(W_{\gamma p},t=0)}{dt}
\left[\frac{x g_A(x,\mu^2)}{Ax g_p(x,\mu^2)}\right]^2 \Phi_A(t_{\rm min})
\,,
\label{eq:charm_A}
\end{eqnarray}
where $xg_A(x,\mu^2)$ is the nuclear gluon distribution; $\Phi_A(t_{\rm min})=\int_{-\infty}^{t_{\rm min}} dt |F_A(t)|^2$, 
where $F_A(t)$ is the nuclear form factor;  $t_{\rm min}=-x^2 m_N^2$ is the minimal momentum transfer squared, where 
$m_N$ is the nucleon mass;
$C_A(\mu^2)/C_p(\mu^2)=(1+\eta_A^2)\bar{R}_{g,A}^2/[(1+\eta^2)\bar{R}_g^2] \approx 0.9$, where $\bar{R}_{g,A}$ and $\eta_A$ are the 
skewness and the ratio of the real to the imaginary parts of the $\gamma A \to V A$ scattering amplitude, respectively.

\begin{figure}[t]
\begin{center}
\epsfig{file=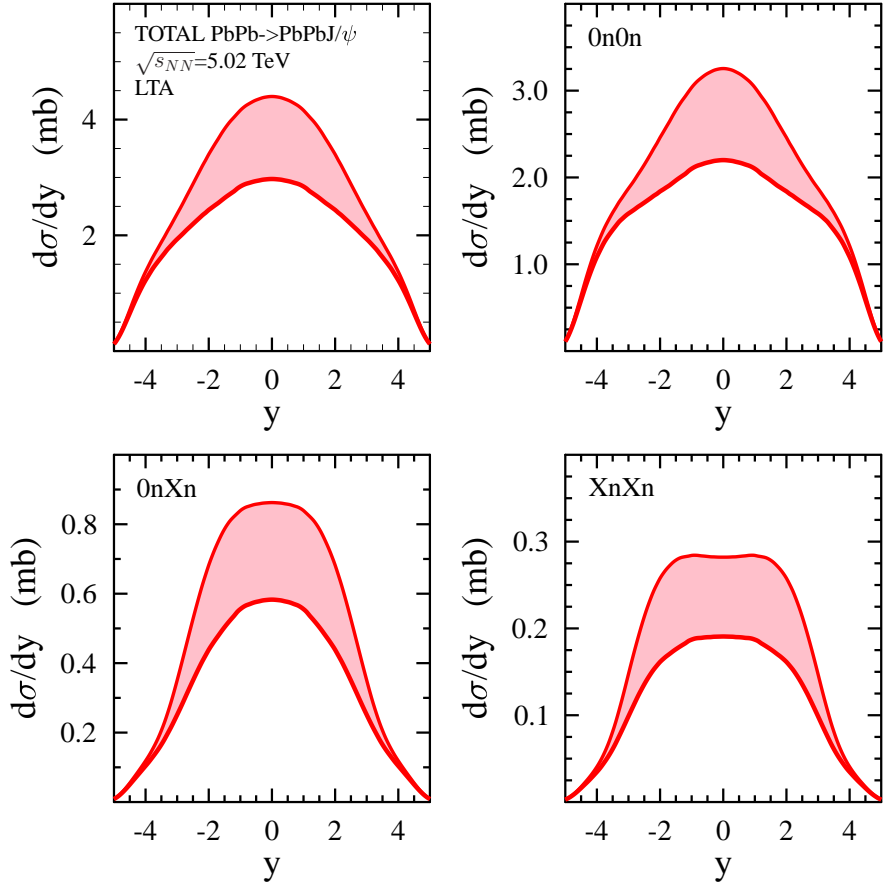,scale=0.6}
 \caption{The $d \sigma_{Pb Pb \to J/\psi Pb Pb}(y)/dy$ cross section as a function of the $J/\psi$ rapidity $y$
 at $\sqrt{s_{NN}}=5.02$ TeV: predictions of LO pQCD with the gluon shadowing ratio $R_g(x,\mu^2)$ of the leading twist nuclear shadowing model.
 The shaded areas span the range of predictions corresponding to the upper and lower limits on $R_g(x,\mu^2)$.
}
 \label{fig:jpsi}
\end{center}
\end{figure}

One can see from Eq.~(\ref{eq:charm_A}) that exclusive photoproduction of charmonia on nuclei directly probes 
the gluon nuclear shadowing quantified by the ratio $R_g(x,\mu^2)=x g_A(x,\mu^2)/[Ax g_p(x,\mu^2)]$. In particular,
a comparison of the nuclear suppression factor extracted from the ALICE data on exclusive $J/\psi$ photoproduction in
Pb-Pb UPCs at $\sqrt{s_{NN}}=2.76$ TeV~\cite{Abbas:2013oua,Abelev:2012ba} to the theoretical predictions based on 
Eq.~(\ref{eq:charm_A}) has given first direct and essentially model-independent evidence of large nuclear gluon shadowing
at $x=0.001$, $R_g(x=10^{-3},\mu^2=3\ {\rm GeV}^2) \approx 0.6$~\cite{Guzey:2013qza,Guzey:2013xba}.
Also, since the values of the resolution scale $\mu^2$ probed in the $J/\psi$ and $\psi(2S)$ cases are close, 
the application of Eq.~(\ref{eq:charm_A}) predicts that the nuclear suppression of the 
$\sigma_{\gamma A \to J/\psi A}(W_{\gamma p})$ and $\sigma_{\gamma A \to \psi(2S) A}(W_{\gamma p})$  cross sections due to gluon nuclear shadowing should also be similar~\cite{Guzey:2014kka}.

\begin{figure}[t]
\begin{center}
\epsfig{file=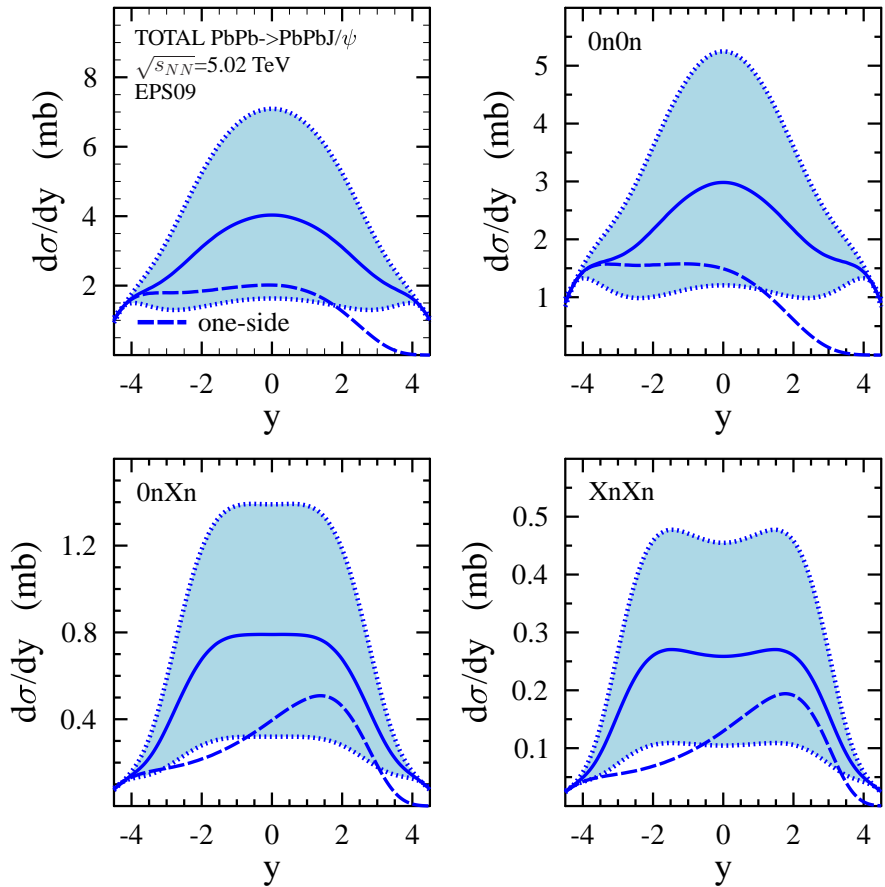,scale=0.6}
 \caption{The $d \sigma_{Pb Pb \to J/\psi Pb Pb}(y)/dy$ cross section as a function of the $J/\psi$ rapidity $y$
 at $\sqrt{s_{NN}}=5.02$ TeV: predictions of LO pQCD with the EPS09 gluon shadowing ratio $R_g(x,\mu^2)$.
 The shaded areas show the uncertainties of $R_g(x,\mu^2)$.
 The dashed curves labeled ``one-side" show the contribution of the first term in Eq.~(\ref{eq:cs_upc}).
}
 \label{fig:jpsi_eps}
\end{center}
\end{figure}

Figures~\ref{fig:jpsi} and \ref{fig:jpsi_eps} present $d \sigma_{Pb Pb \to J/\psi Pb Pb}(y)/dy$ as a function of the 
$J/\psi$ rapidity $y$. They correspond to the calculations using LO pQCD in Eqs.~(\ref{eq:charm}) and (\ref{eq:charm_A}) 
and results of the leading twist nuclear shadowing model~\cite{Frankfurt:2011cs} and the EPS09 nuclear 
parton distribution functions (PDFs)~\cite{Eskola:2009uj}, respectively, 
for the gluon shadowing ratio of $R_g(x,\mu^2)$ at $\mu^2=3$ GeV$^2$.
In Fig.~\ref{fig:jpsi}, the shaded areas span the range of predictions corresponding to the upper and lower limits on $R_g(x,\mu^2)$; in Fig.~\ref{fig:jpsi_eps}, 
the shaded areas show the theoretical uncertainties of $R_g(x,\mu^2)$ in the EPS09 global fit of 
nuclear PDFs.

In Fig.~\ref{fig:jpsi_eps}, we also show the one-side contribution of the first term in Eq.~(\ref{eq:cs_upc}) by the dashed
curves. Similarly to the case of light vector mesons, the one-side rapidity distributions shown in the upper and lower
panels are dramatically different. As we explained above, this happens because of a decrease of the median
impact parameter $b$ in the expression for $N_{\gamma/A}^i(\omega)$~(\ref{eq:N_i}) in the 0nXn and XnXn-channels due to the electromagnetic 
excitations of the nuclei 
leading to an enhanced large-$\omega$ contribution to the photon flux.
Hence, this gives an opportunity to probe the gluon distribution in nuclei $g_A(x,\mu^2)$ in the 0nXn and XnXn-channels
at lower values of $x$ than in the ``total" and 0n0n-channels.

\begin{figure}[t]
\begin{center}
\epsfig{file=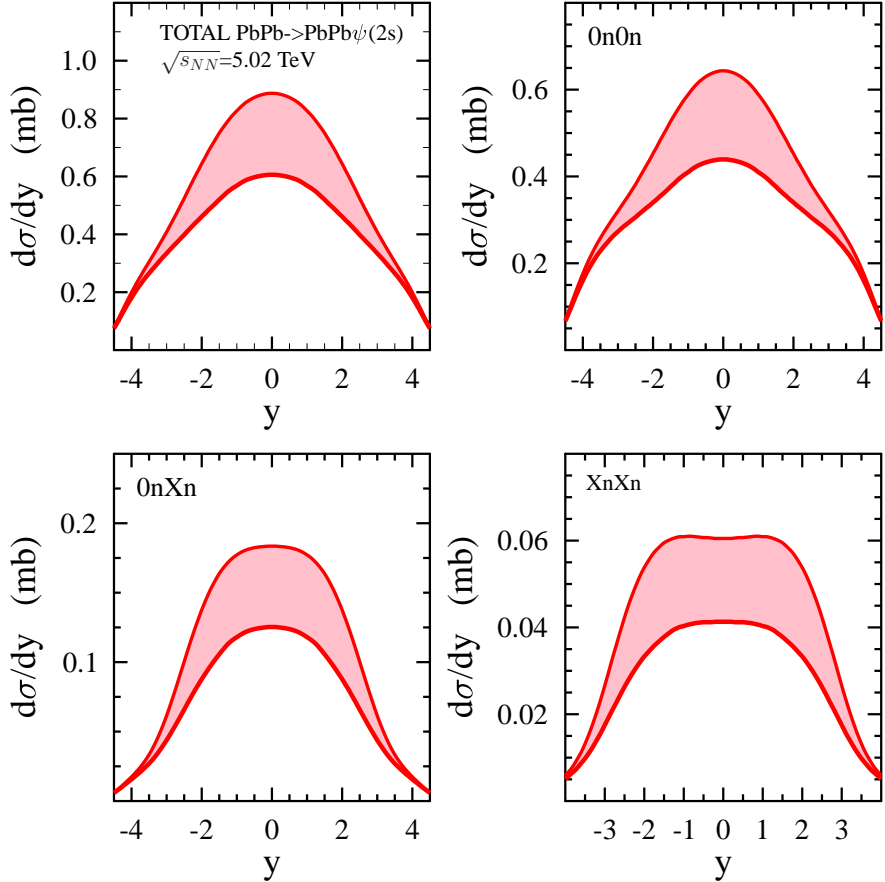,scale=0.6}
 \caption{The $d \sigma_{Pb Pb \to \psi(2S) Pb Pb}(y)/dy$ cross section as a function of the $\psi(2S)$ rapidity $y$
 at $\sqrt{s_{NN}}=5.02$ TeV. See Fig.~\ref{fig:jpsi} for notations.
 }
 \label{fig:psi2}
\end{center}
\end{figure}

Figure~\ref{fig:psi2} presents $d \sigma_{Pb Pb \to \psi(2S) Pb Pb}(y)/dy$ as a function of the 
$\psi(2S)$ rapidity $y$. The shown results
correspond to the LO pQCD calculations with the gluon shadowing ratio $R_g(x,\mu^2)$ at $\mu^2=4$ GeV$^2$
obtained in the leading twist nuclear shadowing model~\cite{Frankfurt:2011cs};
the boundaries of shaded areas correspond to the upper and lower limits on $R_g(x,\mu^2)$.

It is important to stress that since the values of $R_g(x,\mu^2)$ probed in nuclear photoproduction of $J/\psi$ and 
$\psi(2S)$ are close, the suppression of $d\sigma_{Pb Pb \to J/\psi Pb Pb}(y)/dy$ and 
$d\sigma_{Pb Pb \to \psi(2S) Pb Pb}(y)/dy$ due to the leading twist gluon nuclear shadowing
should also be similar. 
Therefore,  the ratio of the $d\sigma_{Pb Pb \to \psi(2S) Pb Pb}(y)/dy$ and $d\sigma_{Pb Pb \to J/\psi Pb Pb}(y)/dy$ cross sections 
is primarily determined by the magnitude and the $W_{\gamma p}$ dependence of the ratio 
of the elementary $\gamma p \to \psi(2S) p$ and $\gamma p \to J/\psi p$ cross sections.
In particular, we find that
\begin{equation}
\frac{d\sigma_{Pb Pb \to \psi(2S) Pb Pb}(y \approx 0)/dy}{d\sigma_{Pb Pb \to J/\psi Pb Pb}(y \approx 0)/dy} =0.17-0.20 \,,
\label{eq:charm_fraction}
\end{equation}
where the lower and upper limits correspond to the calculation of the $\psi(2S)$ cross section at 
$\mu^2 = 3$ GeV$^2$ and $\mu^2 = 4$ GeV$^2$, respectively.
These choices comply with the observation
that energy dependence of the cross section of $\psi(2S)$ photoproduction on the proton is similar to or possibly
somewhat steeper than that for $J/\psi$:  
$R=\sigma_{\gamma p \to \psi(2S) p}/\sigma_{\gamma p \to J/\psi p}=0.166\, (W_{\gamma p}/90 \ {\rm GeV})^{\Delta \delta}$,
where $\Delta \delta=0.24 \pm 0.17$~\cite{Adloff:2002re}.

Note that the result of Eq.~(\ref{eq:charm_fraction})
agrees very well with the predictions of~\cite{Guzey:2014kka} 
at  $\sqrt{s_{NN}}=2.76$ TeV,
$(d\sigma_{Pb Pb \to \psi(2S) Pb Pb}/dy)/(d\sigma_{Pb Pb \to J/\psi Pb Pb}/dy)=0.15-0.16$
for the nuclear gluon shadowing calculated in the leading twist nuclear shadowing model (LTA) and
$(d\sigma_{Pb Pb \to \psi(2S) Pb Pb}/dy)/(d\sigma_{Pb Pb \to J/\psi Pb Pb}/dy)_{|y=0}=0.15^{+0.03}_{-0.01}$
for the calculation with the EPS09 nuclear PDFs,
 as well as with 
 the ratio of
$\psi(2S)$ and $J/\psi$ photoproduction cross sections on the proton measured at HERA~\cite{Adloff:2002re} and 
at the LHC by the LHCb Collaboration~\cite{Aaij:2014iea}.

For the values of the $\psi(2S)$ cross section at $\sqrt{s_{NN}}=5.02$ TeV  and the central rapidity, we predict:
\begin{equation}
\frac{d\sigma_{Pb Pb \to \psi(2S) Pb Pb}(y=0)}{dy} = 0.61-0.89 \ {\rm mb} \,,
\label{eq:charm:abs}
\end{equation}
where the spread in the given values is determined by the theoretical uncertainty of the LTA predictions for the gluon
nuclear shadowing.

It is important to point out that although the approach that we use describes very well the ALICE 
Collaboration data on coherent
$J/\psi$ photoproduction in  Pb-Pb UPCs at $\sqrt{s_{NN}}=2.76$~\cite{Guzey:2013qza,Guzey:2013xba}, its predictions
 for $\psi(2S)$ photoproduction in 
Pb-Pb UPCs at $\sqrt{s_{NN}}=2.76$ TeV and $y=0$~\cite{Guzey:2014kka} 
disagree with the result of the ALICE measurement~\cite{Adam:2015sia}:
the predicted values of $d\sigma_{Pb Pb \to \psi(2S) Pb Pb}(y=0)/dy=0.38-0.56$ mb (LTA) and 
$d\sigma_{Pb Pb \to \psi(2S) Pb Pb}(y=0)/dy=0.42^{+0.24}_{-0.21}$ mb (EPS09)
 significantly underestimate the central experimental value of $d\sigma^{\rm coh}_{\psi(2S)}=0.83 \pm 0.19\ (\rm stat + syst)$ 
 mb (note the large experimental error; the experimental uncertainties for individual channels of the $\psi(2S)$ decay
 are even larger, see Fig.~2 of Ref.~\cite{Adam:2015sia}).
 This translates into the observation that the LTA prediction for the ratio of the $\psi(2S)$ and $J/\psi$ cross
 sections, which weakly depends on $\sqrt{s_{NN}}$, see Eq.~(\ref{eq:charm_fraction}) and its discussion, significantly
 underestimates the experimental value of 
 $d\sigma^{\rm coh}_{\psi(2S)}/dy)/(d\sigma^{\rm coh}_{J/\psi)}/dy = 0.34^{+0.08}_{-0.07} \ (\rm stat + syst)$.
 Note that this disagreement is also present for many other theoretical approaches, see Fig.~5 of Ref.~\cite{Adam:2015sia}.
 Therefore, the measurement of $\psi(2S)$ photoproduction in Pb-Pb UPCs at $\sqrt{s_{NN}}=5.02$ TeV with the statistics 
 (precision) higher than that at  $\sqrt{s_{NN}}=2.76$ TeV should help to better understand the aforementioned disagreement 
 and, thus, constrain the magnitude of the nuclear shadowing suppression in the $\psi(2S)$ case.

\subsection{Exclusive photoproduction of $\Upsilon(1S)$ mesons in pQCD}

An examination shows that the application of Eq.~(\ref{eq:charm}) to exclusive photoproduction of $\Upsilon$ vector mesons 
on the proton fails to reproduce the $W_{\gamma p}$ dependence of the data at leading order (LO) accuracy, whereas providing good description of the data at next-to-leading order (NLO) accuracy.
Indeed, since the gluon distribution of the target is probed at $\mu^2 \approx M_{\Upsilon}^2/4 =22.4$ GeV$^2$,
where the proton LO gluon densities at small $x$ grow approximately as $x g_p (x,\mu^2) \sim 1/x^{\lambda}$ with 
$\lambda \approx 0.4$, Eq.~(\ref{eq:charm}) gives the $W_{\gamma p}$-dependence of the $d\sigma_{\gamma p \to \Upsilon p}(W_{\gamma p}, t=0)/dt$ cross section which is much faster 
than that seen in the data~\cite{Aaij:2015kea,Breitweg:1998ki,Chekanov:2009zz,Abramowicz:2011fa,Adloff:2000vm}.

\begin{figure}[t]
\begin{center}
\epsfig{file=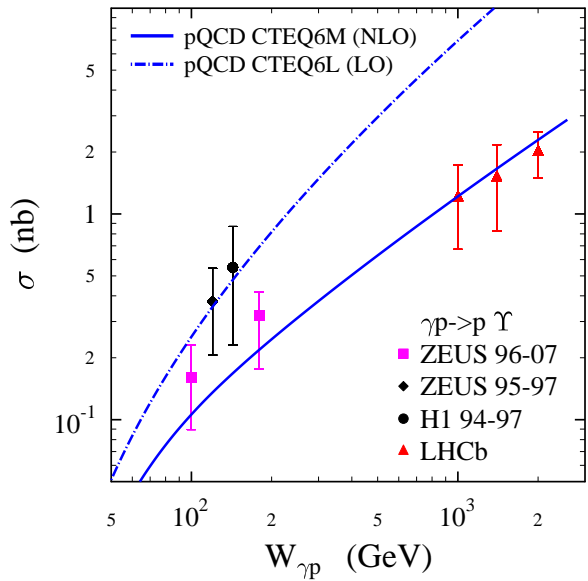,scale=0.45}
 \caption{The $\sigma_{\gamma p \to \Upsilon p}(W_{\gamma p})$ cross section as a function of $W_{\gamma p}$.
 The high-energy ZEUS CollaboratCollaborationion~\cite{Breitweg:1998ki,Chekanov:2009zz,Abramowicz:2011fa}, H1 Collaboration~\cite{Adloff:2000vm} and LHCb Collaboration~\cite{Aaij:2015kea} data are compared to pQCD predictions of Eq.~(\ref{eq:charm}) using the leading-order (CTEQ6L) and next-to-leading order (CTEQ6M) gluon distributions of the proton. }
 \label{fig:Upsilon_proton}
\end{center}
\end{figure}

This is illustrated in Fig.~\ref{fig:Upsilon_proton}, which shows a comparison of the available high-energy data on the 
$t$-integrated cross section $\sigma_{\gamma p \to \Upsilon p}(W_{\gamma p})$ to the LO (dot-dashed curve) and NLO (solid curve)
pQCD predictions for this cross section using the CTEQ6 gluon distributions of the proton~\cite{Nadolsky:2008zw}.
This conclusion also confirms the observation that the LO gluon density of the proton constrained to describe the data on exclusive $J/\psi$ photoproduction on the proton cannot be consistently 
extrapolated to exclusive $\Upsilon$ photoproduction~\cite{Aaij:2015kea,Jones:2013pga}. 

At the same time, the use in Eq.~(\ref{eq:charm}) of the NLO gluon distribution~\cite{Jones:2013pga} obtained by fitting to
the available combined HERA and LHCb data on exclusive $J/\psi$ photoproduction on the proton provides a good description of 
$\Upsilon(1S)$ photoproduction in $pp$ UPCs at $\sqrt{s_{NN}}=7$ and 8 TeV measured by the LHCb Collaboration~\cite{Aaij:2015kea}.
In addition, we  explicitly checked that the use of other NLO gluon distributions of the proton, e.g., the CTEQ6M 
gluon PDF~\cite{Nadolsky:2008zw}, 
reproduces the $\sigma_{\gamma p \to \Upsilon p}(W_{\gamma p})$
cross section with sufficient accuracy --- see the solid curve in Fig.~\ref{fig:Upsilon_proton}.

To summarize, the consistent description of exclusive photoproduction of $\Upsilon$ on both the proton and the
nucleus targets
requires NLO gluon distributions. Therefore, in our predictions for the $\gamma A \to \Upsilon A$ cross section, we use
Eq.~(\ref{eq:charm_A}), where the gluon shadowing ratio $R_g(x,\mu^2)$ is evaluated at
NLO accuracy.

\begin{figure}[t]
\begin{center}
\epsfig{file=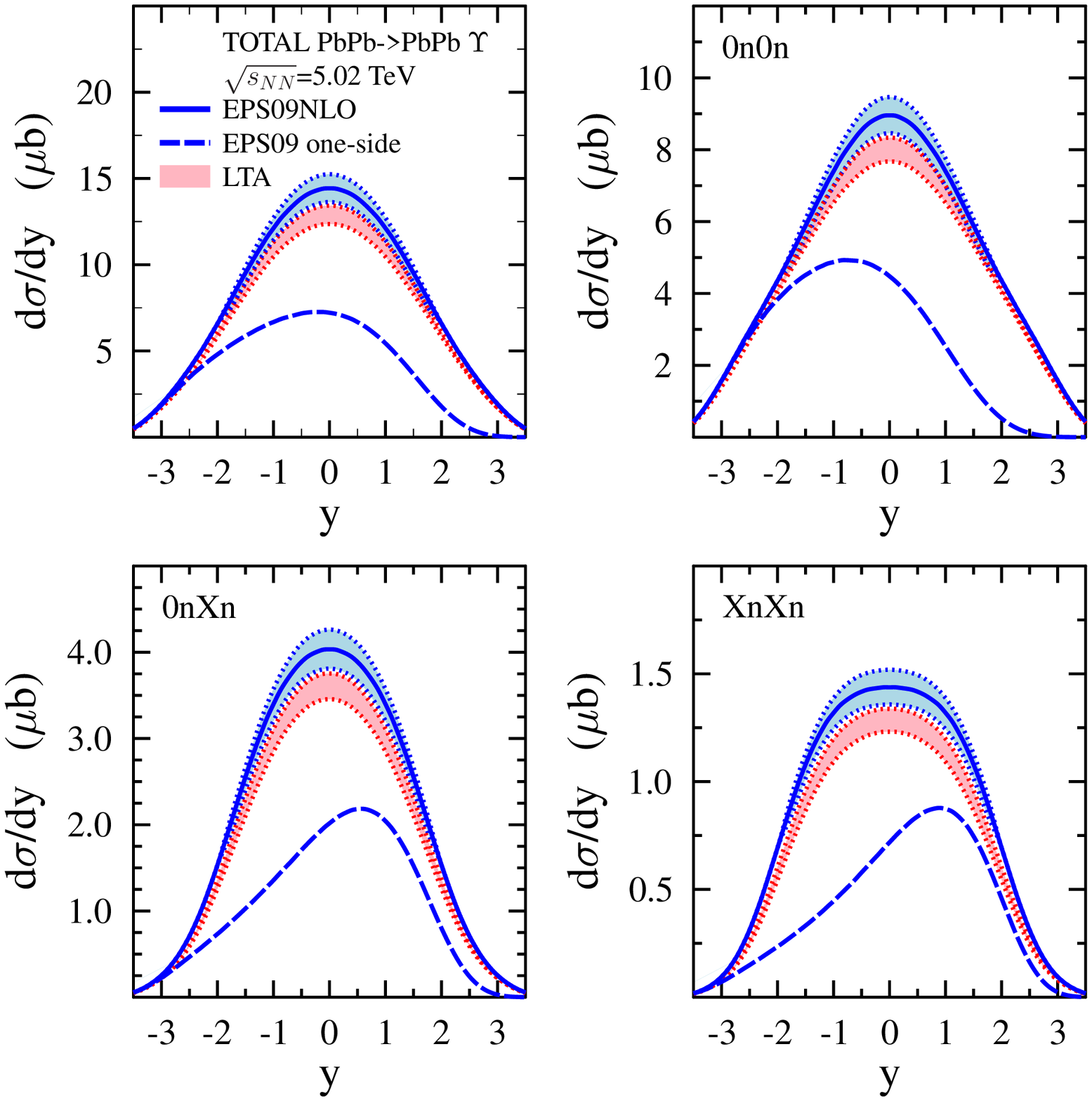,scale=0.6}
 \caption{The $d \sigma_{Pb Pb \to \Upsilon(1S) Pb Pb}(y)/dy$ cross section as a function of the $\Upsilon$ rapidity $y$
 at $\sqrt{s_{NN}}=5.02$ TeV: 
 NLO pQCD calculations with $R_g(x,\mu^2)$ given by the leading twist nuclear
shadowing model (red band labeled ``LTA") and the EPS09 fit (blue solid curve with the band); 
the shaded bands indicate uncertainties of the respective predictions.
The dashed curve labeled ``one-side" is the contribution of the first term in Eq.~(\ref{eq:cs_upc}).
 }
 \label{fig:Upsilon}
\end{center}
\end{figure}

Figure~\ref{fig:Upsilon} shows $d \sigma_{Pb Pb \to \Upsilon(1S) Pb Pb}(y)/dy$ 
as a function of the $\Upsilon(1S)$ rapidity $y$. 
The results correspond to NLO pQCD calculations using for $R_g(x,\mu^2)$ predictions of the leading twist nuclear
shadowing model~\cite{Frankfurt:2011cs} (red band labeled ``LTA") and the EPS09 nuclear PDFs~\cite{Eskola:2009uj} (blue solid curve with the band); 
the shaded bands indicate uncertainties of the respective predictions.
For the calculation with the EPS09 nuclear PDFs, we also show the contribution of the first term in Eq.~(\ref{eq:cs_upc})
(labeled ``one-side") by dashed curves.

One can see from Fig.~\ref{fig:Upsilon} that the one-side cross section at large $y$ gives the dominant contribution
to the rapidity distribution in the 0nXn and XnXn channels. This allows one to be sensitive to the values of 
$W_{\gamma p}$ which are significantly larger than those in the corresponding UPC measurements without the forward
neutron tagging. Hence, it gives an opportunity to probe the nuclear gluon distribution $g_A(x,\mu^2)$ at much lower values of 
$x$.

\section{Discussion and conclusions}
\label{sec:conclusions}

Coherent photoproduction of light and heavy vector mesons in Pb-Pb UPCs at the 
LHC has also been considered in the framework of the color dipole framework~\cite{Goncalves:2011vf,Santos:2014vwa,Lappi:2013am,Lappi:2014eia,Santos:2014zna}. In general, whereas certain combinations of prescriptions for the dipole cross section and
the vector meson light-cone wave functions provide good description of the available data on these processes, 
one has to admit the significant theoretical uncertainty of these predictions due to the choice of the 
dipole cross section at small and large dipole sizes, the value of the quark mass in the photon wave function and the dipole cross section, 
and the shape of the vector meson wave function. Specifically, the cross sections of light $\rho$ and
$\phi$ vector meson photoproduction involve a significant and model-dependent contribution of large-size dipoles,
which in some cases leads to the large nuclear shadowing suppression appropriate to describe 
$\rho$ photoproduction in UPCs at RHIC and the LHC~\cite{Santos:2014vwa}. 
At the same time, the predicted bell-shaped rapidity dependence of Ref.~\cite{Santos:2014vwa} differs from the more 
complicated $y$ dependence that we find. 

In the heavy vector meson case, the photoproduction cross section is dominated by small-size dipoles, 
which leads to the nuclear shadowing suppression due to the elastic interaction of the dipoles with the
target nucleons, which is generally insufficient to describe the data. The description might be improved 
by rescaling the dipole cross section by the factor of $R_g(x,\mu^2)$ quantifying the gluon shadowing; this 
prescription to some extent mimics the large nuclear gluon shadowing in our case, where it is driven
by the inelastic (diffractive) scattering on the target nucleons.   
Thus, studies of photoproduction of heavy vector mesons in heavy ion UPCs provide important constraints on 
the mechanism of nuclear shadowing in hard processes with nuclei.

%
An approach similar to ours was used to make predictions for the rapidity distribution of heavy vector mesons
in proton--nucleus and nucleus--nucleus UPCs at the LHC in Ref.~\cite{Thomas:2016oms}. It was shown that the predictions 
strongly depend on the used parametrization of nuclear PDFs.
Note that our earlier analysis~\cite{Guzey:2013qza} of the data on 
coherent $J/\psi$ photoproduction in $pp$ and $AA$ UPCs at the LHC  showed that a combined pQCD description
of the $\gamma p \to J/\psi p$ and $\gamma A \to J/\psi A$ processes allows us to significantly reduce the ambiguity in the choice of
proton and nucleus PDFs.

The different approaches used in this paper emphasize different aspects of the dynamics of light and heavy vector meson photoproduction on nuclei at high energies. The results presented in this paper attempt to emphasize 
several qualitative features of these processes, which can be checked in Pb-Pb UPCs at the LHC in run 2.
In particular, inelastic nuclear shadowing significantly reduces the magnitude and changes the shape of 
the rapidity distribution of $\rho$ and $\phi$ photoproduction; the large 
leading twist nuclear gluon shadowing suppresses significantly and similarly the $J/\psi$ and $\psi(2S)$ photoproduction 
cross sections; the ratio of these cross sections is determined primarily by the magnitude and the $W_{\gamma p}$ dependence
of the ratio of the elementary $\gamma p \to \psi(2S) p$ and $\gamma p \to J/\psi p$ cross sections;
photoproduction of $\Upsilon(1S)$ allows one to study the hard scale dependence of the leading twist nuclear gluon shadowing.

We also showed that exclusive photoproduction of vector mesons in $AA$ UPCs accompanied by electromagnetic excitations of the ions
followed by the forward neutron emission enhances the contribution of large-energy photons to these processes and allows one
to access the $\gamma A \to VA$ cross section at much larger values of $W_{\gamma p}$.

\acknowledgments

We would like to thank M.~Strikman for useful discussions.


\begin{thebibliography}{99}

\bibitem{Baltz:2007kq} 
  A.~J.~Baltz {\it et al.},
  Phys.\ Rept.\  {\bf 458}, 1 (2008).
  
\bibitem{Adam:2015gsa} 
  J.~Adam {\it et al.} [ALICE Collaboration],
  JHEP {\bf 1509}, 095 (2015).
  
\bibitem{Adler:2002sc} 
  C.~Adler {\it et al.} [STAR Collaboration],
  Phys.\ Rev.\ Lett.\  {\bf 89}, 272302 (2002).

\bibitem{Abelev:2007nb} 
  B.~I.~Abelev {\it et al.} [STAR Collaboration],
  Phys.\ Rev.\ C {\bf 77}, 034910 (2008).

\bibitem{Agakishiev:2011me} 
  G.~Agakishiev {\it et al.} [STAR Collaboration],
  Phys.\ Rev.\ C {\bf 85}, 014910 (2012).
  
\bibitem{Frankfurt:2015cwa} 
  L.~Frankfurt, V.~Guzey, M.~Strikman and M.~Zhalov,
  Phys.\ Lett.\ B {\bf 752}, 51 (2016).
  
\bibitem{Aaij:2013jxj} 
  R.~Aaij {\it et al.} [LHCb Collaboration],
  J.\ Phys.\ G {\bf 40}, 045001 (2013).
  
\bibitem{Aaij:2014iea} 
  R.~Aaij {\it et al.} [LHCb Collaboration],
  J.\ Phys.\ G {\bf 41}, 055002 (2014).

\bibitem{TheALICE:2014dwa} 
  B.~B.~Abelev {\it et al.} [ALICE Collaboration],
  Phys.\ Rev.\ Lett.\  {\bf 113}, no. 23, 232504 (2014).
  
\bibitem{Abbas:2013oua} 
  E.~Abbas {\it et al.} [ALICE Collaboration],
  Eur.\ Phys.\ J.\ C {\bf 73}, no. 11, 2617 (2013).
  
\bibitem{Abelev:2012ba} 
  B.~Abelev {\it et al.} [ALICE Collaboration],
  Phys.\ Lett.\ B {\bf 718}, 1273 (2013).
 
\bibitem{Adam:2015sia} 
  J.~Adam {\it et al.} [ALICE Collaboration],
  Phys.\ Lett.\ B {\bf 751}, 358 (2015).
 
\bibitem{Aaij:2015kea} 
  R.~Aaij {\it et al.} [LHCb Collaboration],
  JHEP {\bf 1509}, 084 (2015).
  
\bibitem{Ryskin:1992ui} 
  M.~G.~Ryskin,
  Z.\ Phys.\ C {\bf 57}, 89 (1993).
  
\bibitem{Jones:2013pga} 
  S.~P.~Jones, A.~D.~Martin, M.~G.~Ryskin and T.~Teubner,
  JHEP {\bf 1311}, 085 (2013).

\bibitem{Guzey:2013qza} 
  V.~Guzey and M.~Zhalov,
  JHEP {\bf 1310}, 207 (2013).
   
\bibitem{Adeluyi:2012ph} 
  A.~Adeluyi and C.~A.~Bertulani,
  Phys.\ Rev.\ C {\bf 85}, 044904 (2012).
  
\bibitem{Guzey:2013xba} 
  V.~Guzey, E.~Kryshen, M.~Strikman and M.~Zhalov,
  Phys.\ Lett.\ B {\bf 726}, 290 (2013).
  
\bibitem{Goncalves:2004bp} 
  V.~P.~Goncalves and M.~V.~T.~Machado,
  Eur.\ Phys.\ J.\ C {\bf 38}, 319 (2004).
  
\bibitem{Lappi:2013am} 
  T.~Lappi and H.~Mantysaari,
  Phys.\ Rev.\ C {\bf 87}, no. 3, 032201 (2013).
  
\bibitem{Goncalves:2014wna} 
  V.~P.~Goncalves, B.~D.~Moreira and F.~S.~Navarra,
  Phys.\ Rev.\ C {\bf 90}, no. 1, 015203 (2014).
 
  
\bibitem{Klein:1999qj} 
  S.~Klein and J.~Nystrand,
  Phys.\ Rev.\ C {\bf 60}, 014903 (1999).
 
\bibitem{Budnev:1974de} 
  V.~M.~Budnev, I.~F.~Ginzburg, G.~V.~Meledin and V.~G.~Serbo,
  Phys.\ Rept.\  {\bf 15}, 181 (1975).

\bibitem{Bertulani:1987tz} 
  C.~A.~Bertulani and G.~Baur,
  Phys.\ Rept.\  {\bf 163}, 299 (1988).
 
 
\bibitem{Glauber:1970jm} 
  R.~J.~Glauber and G.~Matthiae,
  Nucl.\ Phys.\ B {\bf 21}, 135 (1970).

\bibitem{Agashe:2014kda} 
  K.~A.~Olive {\it et al.} [Particle Data Group Collaboration],
  Chin.\ Phys.\ C {\bf 38}, 090001 (2014).

\bibitem{Beiner:1974gc}
  M.~Beiner, H.~Flocard, N.~van Giai and P.~Quentin,
  Nucl.\ Phys.\ A {\bf 238} (1975) 29.

\bibitem{Vidovic:1992ik} 
  M.~Vidovic, M.~Greiner, C.~Best and G.~Soff,
  Phys.\ Rev.\ C {\bf 47}, 2308 (1993).
 
\bibitem{Baltz:2002pp} 
  A.~J.~Baltz, S.~R.~Klein and J.~Nystrand,
  Phys.\ Rev.\ Lett.\  {\bf 89}, 012301 (2002).
  
\bibitem{Bauer:1977iq} 
  T.~H.~Bauer, R.~D.~Spital, D.~R.~Yennie and F.~M.~Pipkin,
  Rev.\ Mod.\ Phys.\  {\bf 50}, 261 (1978)
  [Rev.\ Mod.\ Phys.\  {\bf 51}, 407 (1979)].

  
\bibitem{Frankfurt:2002wc} 
  L.~Frankfurt, M.~Strikman and M.~Zhalov,
  Phys.\ Lett.\ B {\bf 537}, 51 (2002).

\bibitem{Frankfurt:2002sv} 
  L.~Frankfurt, M.~Strikman and M.~Zhalov,
  Phys.\ Rev.\ C {\bf 67}, 034901 (2003).
 
\bibitem{Blaettel:1993ah} 
  B.~Blaettel, G.~Baym, L.~L.~Frankfurt, H.~Heiselberg and M.~Strikman,
  Phys.\ Rev.\ D {\bf 47}, 2761 (1993).
 
  
\bibitem{Donnachie:1994zb} 
  A.~Donnachie and P.~V.~Landshoff,
  Phys.\ Lett.\ B {\bf 348}, 213 (1995).
   
\bibitem{Egloff:1979mg} 
  R.~M.~Egloff {\it et al.},
 Phys.\ Rev.\ Lett.\  {\bf 43}, 657 (1979).
    
\bibitem{Busenitz:1989gq} 
  J.~Busenitz {\it et al.},
  Phys.\ Rev.\ D {\bf 40}, 1 (1989).
 
\bibitem{Derrick:1996af} 
  M.~Derrick {\it et al.} [ZEUS Collaboration],
  Phys.\ Lett.\ B {\bf 377}, 259 (1996).
 
\bibitem{Berger_thesis}
N.~E.~Berger, {\it Measurement of diffractive $\phi$ meson photoproduction at HERA with the H1 fast track trigger}. Ph.D.~thesis, Zurich, ETH, 2006. {\tt http://www-h1.desy.de/publications/theses\_list.html\#YEAR2007}.

\bibitem{Chapin:1985mf} 
  T.~J.~Chapin {\it et al.},
  Phys.\ Rev.\ D {\bf 31}, 17 (1985).

\bibitem{Cool:1981uw} 
  R.~L.~Cool, K.~A.~Goulianos, S.~L.~Segler, H.~Sticker and S.~N.~White,
  Phys.\ Rev.\ Lett.\  {\bf 47}, 701 (1981)
  [Phys.\ Rev.\ Lett.\  {\bf 48}, 61 (1982)].


\bibitem{Ryskin:1995hz} 
  M.~G.~Ryskin, R.~G.~Roberts, A.~D.~Martin and E.~M.~Levin,
  Z.\ Phys.\ C {\bf 76}, 231 (1997).

\bibitem{Martin:2007sb} 
  A.~D.~Martin, C.~Nockles, M.~G.~Ryskin and T.~Teubner,
  Phys.\ Lett.\ B {\bf 662}, 252 (2008).

\bibitem{Jones:2013eda} 
  S.~P.~Jones, A.~D.~Martin, M.~G.~Ryskin and T.~Teubner,
  J.\ Phys.\ G {\bf 41}, 055009 (2014).

\bibitem{Guzey:2014axa} 
  V.~Guzey and M.~Zhalov,
  arXiv:1405.7529 [hep-ph].


\bibitem{Nemchik:1997xb} 
  J.~Nemchik, N.~N.~Nikolaev, E.~Predazzi, B.~G.~Zakharov and V.~R.~Zoller,
  J.\ Exp.\ Theor.\ Phys.\  {\bf 86}, 1054 (1998)
  [Zh.\ Eksp.\ Teor.\ Fiz.\  {\bf 113}, 1930 (1998)].

\bibitem{Hufner:2000jb} 
  J.~Hufner, Y.~P.~Ivanov, B.~Z.~Kopeliovich and A.~V.~Tarasov,
  Phys.\ Rev.\ D {\bf 62}, 094022 (2000).

\bibitem{Ducati:2013bya} 
  M.~B.~G.~Ducati, M.~T.~Griep and M.~V.~T.~Machado,
  Phys.\ Rev.\ C {\bf 88}, 014910 (2013).

\bibitem{Adloff:2002re} 
  C.~Adloff {\it et al.} [H1 Collaboration],
  Phys.\ Lett.\ B {\bf 541}, 251 (2002).


\bibitem{Guzey:2014kka} 
  V.~Guzey and M.~Zhalov,
  arXiv:1404.6101 [hep-ph].
 
\bibitem{Frankfurt:2011cs} 
  L.~Frankfurt, V.~Guzey and M.~Strikman,
  Phys.\ Rept.\  {\bf 512}, 255 (2012).
  
\bibitem{Eskola:2009uj} 
  K.~J.~Eskola, H.~Paukkunen and C.~A.~Salgado,
  JHEP {\bf 0904}, 065 (2009).
  
  
\bibitem{Breitweg:1998ki} 
  J.~Breitweg {\it et al.} [ZEUS Collaboration],
  Phys.\ Lett.\ B {\bf 437}, 432 (1998).

 
\bibitem{Chekanov:2009zz} 
  S.~Chekanov {\it et al.} [ZEUS Collaboration],
  Phys.\ Lett.\ B {\bf 680}, 4 (2009).
  
\bibitem{Abramowicz:2011fa} 
  H.~Abramowicz {\it et al.} [ZEUS Collaboration],
  Phys.\ Lett.\ B {\bf 708}, 14 (2012).
  
\bibitem{Adloff:2000vm} 
  C.~Adloff {\it et al.} [H1 Collaboration],
  Phys.\ Lett.\ B {\bf 483}, 23 (2000).
  
\bibitem{Nadolsky:2008zw} 
  P.~M.~Nadolsky, H.~L.~Lai, Q.~H.~Cao, J.~Huston, J.~Pumplin, D.~Stump, W.~K.~Tung and C.-P.~Yuan,
  Phys.\ Rev.\ D {\bf 78}, 013004 (2008).

\bibitem{Goncalves:2011vf} 
  V.~P.~Goncalves and M.~V.~T.~Machado,
  Phys.\ Rev.\ C {\bf 84}, 011902 (2011).

\bibitem{Santos:2014vwa} 
  G.~Sampaio dos Santos and M.~V.~T.~Machado,
  Phys.\ Rev.\ C {\bf 91}, no. 2, 025203 (2015).
 

\bibitem{Lappi:2014eia} 
  T.~Lappi and H.~MŠntysaari,
  PoS DIS {\bf 2014}, 069 (2014).


\bibitem{Santos:2014zna} 
  G.~Sampaio dos Santos and M.~V.~T.~Machado,
  J.\ Phys.\ G {\bf 42}, no. 10, 105001 (2015).

\bibitem{Thomas:2016oms} 
  J.~Thomas, C.~A.~Bertulani, N.~Brady, D.~B.~Clark, E.~Godat and F.~Olness,
  arXiv:1603.01919 [hep-ph].
 
  
\end{thebibliography}
\end{document}